\documentclass[a4paper,11pt]{article}
\usepackage{amssymb}
\usepackage{amsfonts}
\usepackage{amsmath}
\usepackage{hyperref}
\usepackage{geometry}
\usepackage{lscape}

\newtheorem{theorem}{Theorem}

\newtheorem{remark}{Remark}

\input{tcilatex}
\begin{document}

\title{\textbf{A semiparametric estimation of copula models based on the
method of moments}}
\author{\textbf{Brahim Brahimi, Abdelhakim Necir\thanks{%
Corresponding author: necirabdelhakim@yahoo.fr}} \\
{\small Laboratory of Applied Mathematics, Mohamed Khider University of
Biskra, }\\
{\small Biskra 07000, Algeria}}
\date{}
\maketitle

\begin{abstract}
\noindent Using the classical estimation method of moments, we propose a new
semiparametric estimation procedure for multi-parameter copula models.
Consistency and asymptotic normality of the obtained estimators are
established. By considering an Archimedean copula model, an extensive
simulation study, comparing these estimators with the pseudo maximum
likelihood, rho-inversion and tau-inversion ones, is carried out. We show
that, with regards to the other methods, the moment based estimation is
quick and simple to use with reasonable bias and root mean squared
error.\medskip

\noindent \textbf{MSC classification:} Primary 62G05; Secondary 62G20.

\noindent \textbf{Keywords:} Archimedean copulas; Asymptotic distribution;
Copula models; Measures of association; Method of moments; Semiparametric
models; Statistical inference; Z-estimator.
\end{abstract}

\section{\textbf{Introduction}}

\noindent Recently, considerable attention has been paid to the problem of
inference about copulas. The monographs of Cherubini \textit{et al.} (2004),
Nelsen (2006) and Joe (1997) summarize to some extent the activities in this
area. Roughly speaking, a copula function is a multivariate distribution
function with uniform margins. It is used as a linking block between the
joint distribution function (df) $F$ of a vector of random variables $%
\mathbf{X}=\left( X_{1},...,X_{d}\right) $ and its marginal df's $%
F_{1},...,F_{d}.$ This probabilistic interpretation of copulas is justified
by the famous Sklar's theorem (Sklar, 1959) which states that, under some
mild conditions, there exists a unique copula function $C,$ such that%
\begin{equation*}
F(x_{1},...,x_{d})=C\left( F_{1}\left( x_{1}\right) ,..,F_{d}\left(
x_{d}\right) \right) .
\end{equation*}

\noindent In other words, the copula $C$ is the joint df of the random
vector $\mathbf{U=}\left( U_{1},...,U_{d}\right) ,$ with $U_{j}=F_{j}\left(
X_{j}\right) .$ That is, for $\mathbf{u=}\left( u_{1},...,u_{d}\right) ,$ we
have%
\begin{equation*}
C\left( \mathbf{u}\right) =F\left( F_{1}^{-1}\left( u_{1}\right)
,...,F_{d}^{-1}\left( u_{d}\right) \right) ,
\end{equation*}

\noindent where $F_{j}^{-1}\left( s\right) :=\inf\left\{ x:F_{j}\left(
x\right) \geq s\right\} $ denotes the generalized inverse function (or the
quantile function) of $F_{j}.$\medskip

\noindent A parametric Archimedean copula model arises for $\mathbf{X}$ when
the copula $C$ belong to a class $\mathcal{C}:=\left\{ C_{\mathbf{\theta }},%
\text{ }\mathbf{\theta }\in \mathbf{\mathcal{O}}\right\} ,$ where $\mathbf{%
\mathcal{O}}$ is an open subset of $\mathbb{R}^{r}$ for some integer $r\geq
1.$ Statistical inference on the dependence parameter $\mathbf{\theta }$ is
one of the main topics in multivariate statistical analysis. Several methods
of copula parameter estimation have been developed, including the methods of
concordance (Oakes, 1982, Genest, 1987), fully maximum likelihood (ML),
pseudo maximum likelihood (PML) (Genest\textit{\ et al.}, 1995), inference
function of margins (IFM) (Joe, 1997, 2005), and minimum distance (MD)
(Tsukahara, 2005).\ The performance of the PML procedure vis-a-vis to the
other methods has been discussed by several authors. For example, the
simulation study carried out by Kim \textit{et al}. (2007) has concluded
that the PML method is conceptually almost the same as the IFM one. It
overcomes its non robustness against misspecification of the marginal
distributions. Moreover, by using the PML method, one would not lose any
important statistical insights that would be gained by applying the IFM. An
advantage of the PML over the IFM is that the former does not require
modeling the marginal distributions explicitly. Therefore, the PML estimator
is better than those of the ML and IFM in most practical situations.
However, in time-consuming point of view the ML, PML, IFM and MD methods
require intensive computations, notably when the copula dimension increases.
Moreover, when using these methods the copula density has to be involved,
therefore a serious inaccuracy at boundary points arises. Several numerical
methods are proposed to solve this problem, but\ they are still inefficient
when dealing with high dimensional copula models, more precisely for $d>2$
(see, Yan, 2007, Section 5).\smallskip

\noindent The aim of this paper is to propose an alternative estimation
method similar to the concordance one, avoiding technical problems caused by
copula density and providing estimators with reasonable time-consuming, bias
and root mean squared error (RMSE). The concordance method, also called the $%
\tau $-inversion and $\rho $-inversion, which are based, respectively, on
Kendall's $\tau $ and Spearman's $\rho $ rank correlation coefficients, used
to estimate parametric copula models with at more two parameters. Indeed,
the $\tau $-inversion and $\rho $-inversion methods use the functional
representations of $\tau $ and\ $\rho $ in terms of the underling copula $C$
(Schmid\textit{\ et al.}, 2010), given by%
\begin{align*}
\tau & =\tau \left( C\right) =\frac{1}{2^{d-1}-1}\left\{ 2^{d}\int_{\left[
0,1\right] ^{d}}C\left( \mathbf{u}\right) dC\left( \mathbf{u}\right) \mathbf{%
-}1\right\} \mathbf{,}\text{ } \\
\rho & =\rho \left( C\right) =\frac{d+1}{2^{d}-\left( d+1\right) }\left\{
2^{d}\int_{\left[ 0,1\right] ^{d}}C\left( \mathbf{u}\right) d\mathbf{u}%
-1\right\} .
\end{align*}%
\noindent More precisely, suppose that copula $C$ is a parametric model,
i.e. $C=C_{\mathbf{\theta }},$ then both $\tau $ and\ $\rho $ become
functions in $\mathbf{\theta }$ as well, that is $\tau =\tau \left( \mathbf{%
\theta }\right) $ and $\rho =\rho \left( \mathbf{\theta }\right) .$ Let $%
\widehat{\tau }$ and $\widehat{\rho }$ be, respectively, empirical versions
of $\tau $ and $\rho $ pertaining to the sample $\left( \mathbf{X}_{1},...,%
\mathbf{X}_{n}\right) $ from the random vector $\mathbf{X}$ and suppose that 
$C_{\mathbf{\theta }}$ is one-parameter copula model $($i.e. $r=1).$\ Then,
the estimators of $\theta $ obtained by $\tau $-inversion or $\rho $%
-inversion\ methods are defined by $\widehat{\theta }:=\tau ^{-1}\left( 
\widehat{\tau }\right) $ or $\widehat{\theta }:=\rho ^{-1}\left( \widehat{%
\rho }\right) ,$ where $\tau ^{-1}$ and $\rho ^{-1}$ are the inverses, if
they exist, of functions $\theta \rightarrow \tau \left( \theta \right) $
and $\theta \rightarrow \rho \left( \theta \right) $ respectively. In the
case when $r=2,$ that is when $\mathbf{\theta }=\left( \theta _{1},\theta
_{2}\right) ,$ we have to use jointly the two inversion methods, called $%
\left( \tau ,\rho \right) $-inversion, to have a system of two equations%
\begin{equation}
\tau \left( \theta _{1},\theta _{2}\right) =\widehat{\tau },\text{ }\rho
\left( \theta _{1},\theta _{2}\right) =\widehat{\rho }.  \label{rho-tau-sys}
\end{equation}%
\noindent The consistency of such estimators is discussed in the Appendix
section \ref{Consi}. In conclusion, when the dimension of parameter $\mathbf{%
\theta }$ equals $r,$ we have to use $r$ measures of association, for
example Blomqvist's beta $\beta ,$ Gini's gamma $\gamma ,$ ... (see, Nelsen,
2006, page, 207) which, in general, is not convenient on the choice of
measures point of view. More precisely, suppose that we are dealing with a
parameter $\mathbf{\theta }=\left( \theta _{1},\theta _{2}\right) $ of a
copula model $C_{\theta },$ then one has the right to ask the following
question: What couple among all measures of association have to be chosen to
get a better estimation for $\mathbf{\theta }$? On the other hand, it is
worth mentioning that often there exist difficulties while using Spearman's
rank correlation coefficient. One such difficulty is when using very large
or very small samples. For example, in the case of very large samples, it is
very time consuming to perform Spearman's coefficient since it requires
ranking of the data of all variables.\ Then we have to look for an
alternative more convenient class of measures providing estimators with nice
properties.\ A solution to this problem may be given by applying the
classical method of moments to random variable (rv) $C\left( \mathbf{U}%
\right) .\ $Indeed, let us define the $k$th-moment $M_{k}\left( C\right) ,$
called \textit{copula moment,} of rv $C\left( \mathbf{U}\right) $ as the
expectation of $\left( C\left( \mathbf{U}\right) \right) ^{k},$ that is%
\begin{equation}
M_{k}\left( C\right) :=\mathbf{E}\left[ \left( C\left( \mathbf{U}\right)
\right) ^{k}\right] =\int_{\left[ 0,1\right] ^{d}}\left( C\left( \mathbf{u}%
\right) \right) ^{k}dC\left( \mathbf{u}\right) ,\text{ }k=1,2,...  \label{MK}
\end{equation}%
Notice that the case $k=1$ corresponds to%
\begin{equation*}
M_{1}\left( C\right) =\mathbf{E}\left[ C\left( \mathbf{U}\right) \right] =%
\dfrac{\left( 2^{d-1}-1\right) \tau +1}{2^{d}}.
\end{equation*}%
\noindent In other words, $M_{k}\left( C\right) $ may be considered as a
generalization of Kendall's rank correlation $\tau .$ To our knowledge, the
method of moments is only used in one-parameter copula models, also known by
the $\tau $-inversion method (see for instance, Tsukahara, 2005).\ Note
that, since $0\leq C\left( \mathbf{u}\right) \leq 1,$\textit{\ }then $%
M_{k}\left( C\right) $ are finite for every integer $k.$ Now we are in
position to present a new estimation method that we call \textit{copula
moment (CM) estimation}.\textit{\ }Suppose that, for unknown parameter $%
\mathbf{\theta \in \mathcal{O\subset }}\mathbb{R}^{r}\mathbf{,}$ we have $%
C=C_{\mathbf{\theta }},$ then $M_{k}\left( C\right) =M_{k}\left( \mathbf{%
\theta }\right) ,$ where%
\begin{equation}
M_{k}\left( \mathbf{\theta }\right) :=\int_{\left[ 0,1\right] ^{d}}\left( C_{%
\mathbf{\theta }}\left( \mathbf{u}\right) \right) ^{k}dC_{\mathbf{\theta }%
}\left( \mathbf{u}\right) ,\text{ }k=1,2,...  \label{MKTETA}
\end{equation}

\noindent From equations $\left( \ref{MKTETA}\right) ,$ we may consider $%
\mathbf{M}:\mathbf{\theta \rightarrow }\left( M_{1}\left( \mathbf{\theta }%
\right) ,...,M_{r}\left( \mathbf{\theta }\right) \right) $ as a mapping from 
$\mathcal{O\subset }\mathbb{R}^{r}$ to $\mathbb{R}^{r},$ that will be used
as a means to estimate the parameter $\mathbf{\theta }.$ More precisely, for
a given sample $\left( \mathbf{X}_{1},...,\mathbf{X}_{n}\right) $ of the
random vector $\mathbf{X},$ let us denote $\widehat{\mathbf{\theta }}^{CM}$
as the estimator of $\mathbf{\theta }$ defined by $\left( M_{k}\right)
_{1\leq k\leq r}.$ That is%
\begin{equation}
\widehat{\mathbf{\theta }}^{CM}:=\mathbf{M}^{-1}\left( \widehat{M}_{1},...,%
\widehat{M}_{r}\right) ,  \label{CM-estimator}
\end{equation}

\noindent where $\widehat{M}_{k}$ is the empirical version of $M_{k}\left(
C\right) $ and $\mathbf{M}^{-1}$ is the inverse of the mapping $\mathbf{M},$
provided that it exists.\ The rest of the paper is organized as follows. In
Section \ref{section2}, we present the main steps of the\ copula moment
estimation procedure and establish the consistency and asymptotic normality
of the proposed estimator. In Section \ref{section3}, an application to
multiparameter Archimedean copula models is given. In Section \ref{section 4}%
, an extensive simulation study is carried out to evaluate and compare the
CM based estimation with the PML and $\left( \tau ,\rho \right) $-inversion
methods. Comments and conclusion are given in Section \ref{section5}. The
proofs are relegated to the Appendix.

\section{\label{section2}\textbf{Copula Moments based estimation}}

\noindent In this section we present a semiparametric estimation procedure
for the copula models based on the CM's $\left( \ref{MKTETA}\right) .$ First
suppose that the underlying copula $C$ belongs to a parametric family $C_{%
\mathbf{\theta }},$ with $\mathbf{\theta }=(\theta _{1},\cdots ,\theta
_{r}), $ and satisfies the \textit{concordance ordering condition }of
copulas (see, Nelsen, 2006, page 135), that is:%
\begin{equation}
\text{for every }\mathbf{\theta }_{1},\mathbf{\theta }_{2}\in \mathcal{O}:%
\text{ }\mathbf{\theta }_{1}\neq \mathbf{\theta }_{2}\Longrightarrow C_{%
\mathbf{\theta }_{1}}\left( >\text{ or }<\right) C_{\mathbf{\theta }_{2}}.
\label{concordance}
\end{equation}

\noindent It is clear that this condition implies the well-known \textit{%
identifiability condition} of copulas:%
\begin{equation*}
\text{for every }\mathbf{\theta }_{1},\mathbf{\theta }_{2}\in \mathcal{O}:%
\text{ }\mathbf{\theta }_{1}\neq \mathbf{\theta }_{2}\Longrightarrow C_{%
\mathbf{\theta }_{1}}\neq C_{\mathbf{\theta }_{2}}.
\end{equation*}

\noindent Identifiability is a natural and even a necessary condition: if
the parameter is not identifiable then consistent estimator cannot exist
(see, e.g., van der Vaart, 1998, page 62).\medskip

\noindent For a given sample $\left( \mathbf{X}_{1},...,\mathbf{X}%
_{n}\right) $ from random vector $\mathbf{X=}\left( X_{1},...,X_{d}\right) ,$
we define the corresponding joint empirical df by%
\begin{equation*}
F_{n}\left( \mathbf{x}\right) =n^{-1}\sum_{i=1}^{n}\mathbf{1}\left\{
X_{1i}\leq x_{1},...,X_{di}\leq x_{d}\right\} ,
\end{equation*}

\noindent with $\mathbf{x:=}\left( x_{1},...,x_{d}\right) ,$ and the
marginal empirical df's pertaining to the sample $\left(
X_{j1},...,X_{jn}\right) ,$ from rv $X_{j},$ by%
\begin{equation}
F_{jn}\left( x_{j}\right) =n^{-1}\sum_{i=1}^{n}\mathbf{1}\left\{ X_{ji}\leq
x_{j}\right\} ,\text{ }j=1,...,d.  \label{Fn}
\end{equation}%
\ \ 

\noindent According to Deheuvels (1979), the empirical copula function is
defined by%
\begin{equation*}
C_{n}\left( \mathbf{u}\right) :=F_{n}\left( F_{1n}^{-1}\left( u_{1}\right)
,...,F_{dn}^{-1}\left( u_{d}\right) \right) ,\text{ for }\mathbf{u}\in\left[
0,1\right] ^{d},
\end{equation*}

\noindent where $F_{jn}^{-1}\left( s\right) :=\inf \left\{ x:F_{jn}\left(
x\right) \geq s\right\} $ denotes the empirical quantile function pertaining
to df $F_{jn}.$ We are now in position to present, in three steps, the
semiparametric CM-based estimation:

\begin{itemize}
\item \textbf{Step 1}: For each $j=1,...,d,$ compute $\widehat{U}%
_{ji}:=F_{jn}\left( X_{ji}\right) ,$ then set 
\begin{equation*}
\widehat{\mathbf{U}}_{i}:=\left( \widehat{U}_{1i},...,\widehat{U}%
_{di}\right) ,\text{ }i=1,...,n.
\end{equation*}

\item \textbf{Step 2}: For each $k=1,...,r,$ compute%
\begin{equation}
\widehat{M}_{k}:=n^{-1}\sum_{i=1}^{n}\left( C_{n}\left( \widehat{\mathbf{U}}%
_{i}\right) \right) ^{k}.  \label{gamacha}
\end{equation}

\noindent as the natural estimators of CM's $M_{k}$ given in equation $%
\left( \ref{MK}\right) .$

\item \textbf{Step 3}\textit{:} Solve the following system%
\begin{equation}
\left\{ 
\begin{array}{l}
M_{1}\left( \theta _{1},...,\theta _{r}\right) =\widehat{M}_{1} \\ 
M_{2}\left( \theta _{1},...,\theta _{r}\right) =\widehat{M}_{2} \\ 
\vdots \\ 
M_{r}\left( \theta _{1},...,\theta _{r}\right) =\widehat{M}_{r}.%
\end{array}%
\ \right.  \label{system}
\end{equation}

The obtained solution $\mathbf{\hat{\theta}}^{CM}:=\left( \widehat{\theta }%
_{1},...,\widehat{\theta }_{r}\right) $ is called the CM estimator for $%
\mathbf{\theta }.\medskip $
\end{itemize}

\noindent Consistency and asymptotic normality of $\mathbf{\hat{\theta}}%
^{CM} $ are stated in Theorem \ref{TH1} below whose proof is relegated to
the Appendix \ref{ProofTH1}. For convenience we set%
\begin{equation}
L_{k}\left( \mathbf{u};\mathbf{\theta }\right) :=\left( C_{\mathbf{\theta }%
}\left( \mathbf{u}\right) \right) ^{k}-M_{k}\left( \mathbf{\theta }\right) 
\text{ and }\mathbf{L}\left( \mathbf{u};\mathbf{\theta }\right) =\left(
L_{1}\left( \mathbf{u};\mathbf{\theta }\right) ,...,L_{r}\left( \mathbf{u};%
\mathbf{\theta }\right) \right) .  \label{Lk}
\end{equation}%
\noindent Let $\mathbf{\theta }_{0}$ be the true value of $\mathbf{\theta }$
and assume that the following assumptions $\left[ H.1\right] -\left[ H.3%
\right] $ hold.

\begin{itemize}
\item $\left[ H.1\right] $ $\mathbf{\theta }_{0}\in \mathcal{O}\subset 
\mathbb{R}^{r}$ is the unique zero of the mapping $\mathbf{\theta }%
\rightarrow \int_{\left[ 0,1\right] ^{d}}\mathbf{L}\left( \mathbf{u};\mathbf{%
\theta }\right) dC_{\mathbf{\theta }_{0}}\left( \mathbf{u}\right) $ which is
defined from $\mathcal{O}$ to $\mathbb{R}^{r}.$

\item $\left[ H.2\right] $ $\mathbf{L}\left( \cdot ;\mathbf{\theta }\right) $
is differentiable with respect to $\mathbf{\theta }$ with the Jacobian
matrix denoted by%
\begin{equation*}
\overset{\bullet }{\mathbf{L}}\left( \mathbf{u};\mathbf{\theta }\right) :=%
\left[ \frac{\partial L_{k}\left( \mathbf{u};\mathbf{\theta }\right) }{%
\partial \theta _{\ell }}\right] _{r\times r},
\end{equation*}%
$\overset{\bullet }{\mathbf{L}}\left( \mathbf{u};\mathbf{\theta }\right) $
is continuous both in $\mathbf{u}$ and $\mathbf{\theta },$ and the Euclidian
norm $\left\vert \overset{\bullet }{\mathbf{L}}\left( \mathbf{u};\mathbf{%
\theta }\right) \right\vert $ is dominated by a $dC_{\mathbf{\theta }}$%
-integrable function $h\left( \mathbf{u}\right) .$

\item $\left[ H.3\right] $ The $r\times r$ matrix $A_{0}:=\int_{\left[ 0,1%
\right] ^{d}}\overset{\bullet }{\mathbf{L}}\left( \mathbf{u};\mathbf{\theta }%
_{0}\right) dC_{\mathbf{\theta }_{0}}\left( \mathbf{u}\right) $ is
nonsingular.$\medskip $
\end{itemize}

\begin{theorem}
\label{TH1}Assume that the \textit{concordance ordering condition} $\left( %
\ref{concordance}\right) $ and assumptions $\left[ H.1\right] -\left[ H.3%
\right] $ hold. Then with probability tending to one as $n\rightarrow \infty
,$ there exists a solution $\widehat{\mathbf{\theta }}^{CM}$ to the system $%
\left( \ref{system}\right) $ which converges to $\mathbf{\theta }_{0}.$
Moreover 
\begin{equation*}
\sqrt{n}\left( \widehat{\mathbf{\theta }}^{CM}-\mathbf{\theta }_{0}\right) 
\overset{\mathcal{D}}{\rightarrow }\mathcal{N}\left( \mathbf{0}%
,A_{0}^{-1}D_{0}\left( A_{0}^{-1}\right) ^{T}\right) ,\text{ as }%
n\rightarrow \infty ,
\end{equation*}%
where $D_{0}:=var\left\{ \mathbf{L}\left( \mathbf{\xi };\mathbf{\theta }%
_{0}\right) +\mathbf{V}\left( \mathbf{\xi };\mathbf{\theta }_{0}\right)
\right\} $ and $\mathbf{V}\left( \mathbf{\xi };\mathbf{\theta }_{0}\right)
=\left( V_{1}\left( \mathbf{\xi };\mathbf{\theta }_{0}\right)
,...,V_{r}\left( \mathbf{\xi };\mathbf{\theta }_{0}\right) \right) $ with 
\begin{equation*}
V_{k}\left( \mathbf{\xi };\mathbf{\theta }_{0}\right) :=\sum_{j=1}^{d}\int_{%
\left[ 0,1\right] ^{d}}\frac{\partial \left( C_{\mathbf{\theta }_{0}}\left( 
\mathbf{u}\right) \right) ^{k}}{\partial u_{j}}\left( \mathbf{1}\left\{ \xi
_{j}\leq u_{j}\right\} -u_{j}\right) dC_{\mathbf{\theta }_{0}}\left( \mathbf{%
u}\right) ,\text{ }k=1,...,r,
\end{equation*}%
where $\mathbf{\xi :=}\left( \xi _{1},...,\xi _{d}\right) $ is a $\left(
0,1\right) ^{d}$-uniform random vector with joint df $C_{\mathbf{\theta }%
_{0}}.$
\end{theorem}

\begin{remark}
The asymptotic variance $A_{0}^{-1}D_{0}\left( A_{0}^{-1}\right) ^{T}$ may
be consistently estimated by the sample variance of of the sequence of rv's $%
\left( \widehat{A}_{i}^{-1}\widehat{D}_{i}\left( \widehat{A}_{i}^{-1}\right)
^{T},\text{ }i=1,...,n\right) $ where 
\begin{equation*}
\widehat{A}_{i}:=\int_{\left[ 0,1\right] ^{d}}\overset{\bullet }{\mathbf{L}}%
\left( \mathbf{u};\widehat{\mathbf{\theta }}^{CM}\right) dC_{\widehat{%
\mathbf{\theta }}^{CM}}\left( \mathbf{u}\right) \text{ and }\widehat{D}_{i}:=%
\mathbf{L}\left( \widehat{\mathbf{U}}_{i};\widehat{\mathbf{\theta }}%
^{CM}\right) +\mathbf{V}\left( \widehat{\mathbf{U}}_{i};\widehat{\mathbf{%
\theta }}^{CM}\right) ,
\end{equation*}%
as is done, in Genest et al. (1995) and Tsukahara (2005) in the case of
PML's estimator and Z-estimator respectively. For more details on the
Z-estimation theory, one refers to van der Vaart (1998), page 41.
\end{remark}

\section{\label{section3}\textbf{Application: Archimedean copula models}}

\noindent As application to the CM estimation method, we consider the
Archimedean copula family defined by $C(\mathbf{u})=\varphi ^{-1}\left(
\sum_{j=1}^{d}\varphi (u_{j})\right) ,$ where $\varphi :\left[ 0,1\right]
\rightarrow \mathbb{R}$ is a twice differentiable function called the
generator, satisfying: $\varphi \left( 1\right) =0,$ $\varphi ^{\prime
}\left( x\right) <0,$ $\varphi ^{\prime \prime }\left( x\right) \geq 0$ for
any $x\in \left( 0,1\right) .$ The notation $\varphi ^{-1}$ stands for the
inverse function of $\varphi .\ $Archimedean copulas are easy to construct
and have nice properties. A variety of known copula families belong to this
class, including the models of Gumbel, Clayton, Frank, ... (see, Table 4.1
in Nelsen, 2006, page 116). Let $\mathbf{K}_{C}(s):=P\left( C\left( \mathbf{U%
}\right) \leq s\right) ,$ $s\in \left[ 0,1\right] ,$ be the df of rv $%
C\left( \mathbf{U}\right) ,$ then equation $\left( \ref{MK}\right) $ may be
rewritten into:%
\begin{equation*}
M_{k}\left( C\right) =\int_{0}^{1}s^{k}d\mathbf{K}_{C}(s),\text{ }k=1,2,....
\end{equation*}

\noindent Suppose now, for unknown $\mathbf{\theta \in \mathcal{O},}$ that $%
\varphi =\varphi _{\mathbf{\theta }},$ it follows that $C=C_{\mathbf{\theta }%
},$ $\mathbf{K}_{C}=\mathbf{K}_{\mathbf{\theta }}$ and $M_{k}\left( C\right)
=M_{k}\left( \mathbf{\theta }\right) ,$ that is%
\begin{equation*}
M_{k}\left( \mathbf{\theta }\right) =\int_{0}^{1}s^{k}d\mathbf{K}_{\mathbf{%
\theta }}(s),\text{ }k=1,2,...,
\end{equation*}

\noindent Notice that, one of the nice properties of Archimedean copula is
that the df $\mathbf{K}_{C}$ of $C\left( \mathbf{U}\right) $ may be
represented in terms of the first and second derivatives of the generator.
Indeed from Theorem 4.3.4 in Nelsen (2006), for any $s\in \left[ 0,1\right]
, $ $\mathbf{K}_{\mathbf{\theta }}(s)=s-\varphi _{\mathbf{\theta }}\left(
s\right) /\varphi _{\mathbf{\theta }}^{\prime }\left( s\right) ,$ it follows
that the corresponding density is $\mathbf{K}_{\mathbf{\theta }}^{\prime
}(s)=\varphi _{\mathbf{\theta }}^{\prime \prime }\left( s\right) \varphi _{%
\mathbf{\theta }}\left( s\right) /\left( \varphi _{\mathbf{\theta }}^{\prime
}\left( s\right) \right) ^{2}.$\ Therefore the $k$th CM, defined in $\left( %
\ref{MK}\right) ,$ may be rewritten into%
\begin{equation}
M_{k}\left( \mathbf{\theta }\right) =\int_{0}^{1}s^{k}\frac{\varphi _{%
\mathbf{\theta }}^{\prime \prime }\left( s\right) \varphi _{\mathbf{\theta }%
}\left( s\right) }{\left( \varphi _{\mathbf{\theta }}^{\prime }\left(
s\right) \right) ^{2}}ds,\text{ }k=1,2,...  \label{lamdateta}
\end{equation}%
In terms of $\mathbf{K}_{\mathbf{\theta }},$ the assumptions $\left[ H.1%
\right] -\left[ H.3\right] $ and Theorem \ref{TH1} may be rephrased,
respectively, to $\left[ H.1^{\prime }\right] -\left[ H.3^{\prime }\right] $
and Theorem \ref{TH2} below. For convenience, we set 
\begin{equation*}
\mathcal{L}\left( t;\mathbf{\theta }\right) \mathbf{=}\left( \mathcal{L}%
_{1}\left( t;\mathbf{\theta }\right) ,...,\mathcal{L}_{r}\left( t;\mathbf{%
\theta }\right) \right) \text{ with }\mathcal{L}_{k}\left( t;\mathbf{\theta }%
\right) :=t^{k}-M_{k}\left( \mathbf{\theta }\right) .
\end{equation*}

\begin{itemize}
\item $\left[ H.1^{\prime }\right] $ $\mathbf{\theta }_{0}\in \mathcal{O}%
\subset \mathbb{R}^{r}$ is the unique zero of the mapping $\mathbf{\theta }%
\rightarrow \int_{0}^{1}\mathcal{L}\left( \mathbf{t};\mathbf{\theta }\right)
d\mathbf{K}_{\mathbf{\theta }_{0}}(t)$ that is defined from $\mathcal{O}$ to 
$\mathbb{R}^{r}.$

\item $\left[ H.2^{\prime }\right] $ $\mathcal{L}\left( \cdot ;\mathbf{%
\theta }\right) $ is differentiable with respect to $\mathbf{\theta }$ with
the Jacobian matrix denoted by%
\begin{equation*}
\overset{\bullet }{\mathcal{L}}\left( t;\mathbf{\theta }\right) :=\left[ 
\frac{\partial M_{k}\left( \mathbf{\theta }\right) }{\partial \theta _{\ell }%
}\right] _{r\times r},
\end{equation*}%
$\overset{\bullet }{\mathcal{L}}\left( t;\mathbf{\theta }\right) $ is
continuous both in $t$ and $\mathbf{\theta },$ and the Euclidian norm $%
\left\vert \overset{\bullet }{\mathcal{L}}\left( t;\mathbf{\theta }\right)
\right\vert $ is dominated by a $d\mathbf{K}_{\mathbf{\theta }}$-integrable
function $h\left( t\right) .$

\item $\left[ H.3^{\prime}\right] $ The $r\times r$ matrix $\mathcal{A}%
_{0}:=\int_{0}^{1}\overset{\bullet}{\mathcal{L}}\left( t;\mathbf{\theta}%
_{0}\right) d\mathbf{K}_{\mathbf{\theta}_{0}}\left( t\right) $ is
nonsingular.
\end{itemize}

\begin{theorem}
\label{TH2}Assume that \textit{concordance ordering condition} $\left( \ref%
{concordance}\right) $ and assumptions $\left[ H.1^{\prime }\right] -\left[
H.3^{\prime }\right] $ hold. Then with probability tending to one as $%
n\rightarrow \infty ,$ there exists a solution $\widehat{\mathbf{\theta }}%
^{CM}$ to the system $\left( \ref{system}\right) $ which converges to $%
\mathbf{\theta }_{0}.$ Moreover 
\begin{equation*}
\sqrt{n}\left( \widehat{\mathbf{\theta }}^{CM}-\mathbf{\theta }_{0}\right) 
\overset{\mathcal{D}}{\rightarrow }\mathcal{N}\left( \mathbf{0},\mathcal{A}%
_{0}^{-1}\mathcal{D}_{0}\left( \mathcal{A}_{0}^{-1}\right) ^{T}\right) ,%
\text{ as }n\rightarrow \infty ,
\end{equation*}%
where%
\begin{equation*}
\mathcal{D}_{0}:=var\left\{ \mathcal{L}\left( \xi ;\mathbf{\theta }%
_{0}\right) +\int_{0}^{1}\mathbf{g}\left( t\right) \left( \mathbf{1}\left\{
\xi \leq t\right\} -t\right) d\mathbf{K}_{\mathbf{\theta }_{0}}\left(
t\right) \right\} ,
\end{equation*}%
where $\xi $ is a $\left( 0,1\right) $-uniform rv and $\mathbf{g}\left(
t\right) \mathbf{:=}\left( kt^{k-1}\right) _{1\leq k\leq r}$ is $r$%
-dimensional vector.
\end{theorem}

\subsection{\textbf{Illustrative example}}

\noindent The Gumbel family is an Archimedean copula defined by%
\begin{equation*}
C_{\beta}(\mathbf{u})=\exp\left( -\left( \sum_{j=1}^{d}\left( -\ln
u_{j}\right) ^{\beta}\right) ^{1/\beta}\right) ,\text{ }\beta\geq1,
\end{equation*}

\noindent with generator $\varphi _{\beta }\left( t\right) =\left( -\ln
t\right) ^{\beta },$ $\beta \geq 1.$ For the sake of flexibility in data
modeling, it is better to use the multi-parameters copula models than the
one-parameter ones. To have a copula with more than one parameter, we use,
for instance, the transformed (or distorted) copula defined by%
\begin{equation*}
C_{\Gamma }\left( \mathbf{u}\right) =\Gamma ^{-1}\left( C\left( \Gamma
\left( u_{1}\right) ,...,\Gamma \left( u_{d}\right) \right) \right) ,
\end{equation*}

\noindent where $\Gamma :\left[ 0,1\right] \rightarrow \left[ 0,1\right] $
is a continuous, concave and strictly increasing function with $\Gamma
\left( 0\right) =0$ and $\Gamma \left( 1\right) =1.$ As an example, suppose
that $\Gamma =\Gamma _{\alpha },$ with $\Gamma _{\alpha }\left( t\right)
=\exp \left( 1-t^{-\alpha }\right) ,$ $\alpha >0$ and consider the Gumbel
copula $C_{\beta },$ then the transformed copula $C_{\alpha ,\beta }\left( 
\mathbf{u}\right) =\Gamma _{\alpha }^{-1}\left( C_{\beta }\left( \Gamma
_{\alpha }\left( u_{1}\right) ,...,\Gamma _{\alpha }\left( u_{d}\right)
\right) \right) $ is given by%
\begin{equation}
C_{\alpha ,\beta }\left( \mathbf{u}\right) :=\left( \left(
\sum_{j=1}^{d}\left( u_{j}^{-\alpha }-1\right) ^{\beta }\right) ^{1/\beta
}+1\right) ^{-1/\alpha },  \label{model}
\end{equation}

\noindent which is also a two-parameter Archimedean copula with generator $%
\varphi _{\alpha ,\beta }\left( t\right) :=\left( t^{-\alpha }-1\right)
^{\beta }.$ Note that $C_{\alpha ,\beta }$ verifies the concordance ordering
condition $\left( \ref{concordance}\right) $\textbf{\ (}see,\textbf{\ }%
Nelsen, 2006, page, 145\textbf{)}. By an elementary calculation we get the $%
k $th CM:%
\begin{equation*}
M_{k}\left( \alpha ,\beta \right) =\frac{(k+1)\beta +\alpha \beta -k}{\left(
k+1\right) ^{2}\beta +\left( k+1\right) \alpha \beta }.
\end{equation*}%
\noindent In particular the first two CM's are%
\begin{equation*}
M_{1}\left( \alpha ,\beta \right) :=\dfrac{2\beta +\alpha \beta -1}{4\beta
+2\alpha \beta }\text{ and }M_{2}\left( \alpha ,\beta \right) :=\dfrac{%
3\beta +\alpha \beta -2}{9\beta +3\alpha \beta }.
\end{equation*}

\noindent Let $\left( \mathbf{X}_{1},...,\mathbf{X}_{n}\right) $ be a sample
of random vector $\mathbf{X}=\left( X_{1},...,X_{d}\right) ,$ then the CM
estimator $\left( \widehat{\alpha},\widehat{\beta}\right) $ of $\left(
\alpha,\beta\right) $ is the unique solution of the system%
\begin{equation*}
\left\{ 
\begin{array}{c}
M_{1}\left( \alpha,\beta\right) =\widehat{M}_{1} \\ 
M_{2}\left( \alpha,\beta\right) =\widehat{M}_{2}.%
\end{array}
\right.
\end{equation*}
\noindent That is%
\begin{equation}
\widehat{\alpha}=\frac{8\widehat{M}_{1}-9\widehat{M}_{2}-1}{1-4\widehat{M}%
_{1}+3\widehat{M}_{2}},\text{ }\widehat{\beta}=\frac{1-4\widehat{M}_{1}+3%
\widehat{M}_{2}}{\left( 1-2\widehat{M}_{1}\right) \left( 1-3\widehat{M}%
_{2}\right) }.  \label{sol}
\end{equation}

\section{\label{section 4}\textbf{Simulation study}}

\noindent First notice that all numerical computations are performed on a
personal computer with a microprocessor speed of 2.4 GHz. To evaluate and
compare the performance of CM's estimator with the PML and $\left( \tau
,\rho \right) $-inversion estimators, a simulation study is carried out by
considering the transformed bivariate Gumbel copula family $C_{\alpha ,\beta
}$ defined above.\ The evaluation of the performance is based on the bias
and the RMSE defined as follows:{\small {%
\begin{equation}
\text{Bias}=\frac{1}{N}\sum_{i=1}^{N}\hat{\theta}_{i}-\theta ,\ \text{RMSE}%
=\left( \frac{1}{N}\sum_{i=1}^{N}\left( \hat{\theta}_{i}-\theta \right)
^{2}\right) ^{1/2},  \label{b}
\end{equation}%
}}where $\hat{\theta}_{i}$ is an estimator (from the considered method) of $%
\theta $ from the $i$th samples for $N$ generated samples from the
underlying copula. In both parts, we selected $N=1000.$ The procedure
outlined in Section \ref{section2} is repeated for different sample sizes $n$
with $n=30,50,100,200$ to assess the improvement in the bias and RMSE of the
estimators with increasing sample size. Furthermore, the simulation
procedure is repeated for a large set of parameters of the true copula $%
C_{\alpha ,\beta }.$ For each sample, by using formulas $\left( \ref{sol}%
\right) ,$ we obtain the CM-estimator $\left( \widehat{\alpha }_{i},\widehat{%
\beta }_{i}\right) $ of $\left( \alpha ,\beta \right) $ for $i=1,...,N,$ and
the estimators $\widehat{\alpha }$ and $\widehat{\beta }$ are given by $%
\widehat{\alpha }=\frac{1}{N}\sum_{i=1}^{N}\widehat{\alpha }_{i}$ and $%
\widehat{\beta }=\frac{1}{N}\sum_{i=1}^{N}\widehat{\beta }_{i}.$ The choice
of the true values of the parameter $\left( \alpha ,\beta \right) $ have to
be meaningful, in the sense that each couple of parameters assigns a value
of one of the dependence measure, that is weak, moderate and strong
dependence.\ In other words, if we consider Kendall's $\tau $ as a
dependence measure, then we should select values for copula parameters that
correspond to specified\ values of $\tau $ by means of the equation 
\begin{equation}
\tau \left( \alpha ,\beta \right) =4\int_{\left[ 0,1\right] ^{2}}C_{\alpha
,\beta }\left( u_{1},u_{2}\right) dC_{\alpha ,\beta }\left(
u_{1},u_{2}\right) \mathbf{-}1.  \label{tau}
\end{equation}%
The selected values of the true parameters are summarized in Table 1: 
\begin{landscape}
\tiny{
\begin{table}[h] \centering%
$%
\begin{tabular}{c|c|c}
${\small \tau }$ & ${\small \alpha }$ & ${\small \beta }$ \\ \hline\hline
${\small 0.01}$ & ${\small 0.1}$ & ${\small 1.059}$ \\ \hline
${\small 0.2}$ & ${\small 0.2}$ & ${\small 1.137}$ \\ \hline
${\small 0.5}$ & ${\small 0.5}$ & ${\small 1.600}$ \\ \hline
${\small 0.8}$ & ${\small 0.9}$ & ${\small 3.450}$ \\ \hline
\end{tabular}%
\ \ \ \ \ $%
\caption{The true parameters of transformed Gumbel copula used for the
simulation study.}\label{T1}%
\end{table}%

\noindent%
\begin{table}[h] \centering%
$%
\begin{tabular}{ccccc|cccc|ccccl}
& \multicolumn{4}{c|}{${\small \tau=0.01}$} & \multicolumn{4}{|c|}{${\small %
\tau=0.5}$} & \multicolumn{4}{|c}{${\small \tau =0.8}$} &  \\ \hline
& \multicolumn{2}{c}{${\small \alpha=0.1}$} & \multicolumn{2}{c|}{${\small %
\beta=1.059}$} & \multicolumn{2}{|c}{${\small \alpha=0.5}$} &
\multicolumn{2}{c|}{${\small \beta=1.6}$} & \multicolumn{2}{|c}{${\small %
\alpha=0.9}$} & \multicolumn{2}{c}{${\small \beta =3.45}$} &  \\ \hline
\multicolumn{1}{l}{${\small n}$} & {\small Bias} & {\small RMSE} & {\small %
Bias} & {\small RMSE} & {\small Bias} & {\small RMSE} & {\small Bias} &
{\small RMSE} & {\small Bias} & {\small RMSE} & {\small Bias} & {\small RMSE}
& {\small CPU} \\ \hline\hline
\multicolumn{1}{l}{${\small 30}$} & \multicolumn{1}{r}{${\small -0.081}$} &
\multicolumn{1}{r}{${\small 0.330}$} & \multicolumn{1}{r}{${\small 0.032}$}
& \multicolumn{1}{r|}{${\small 0.180}$} & \multicolumn{1}{|r}{${\small -0.051%
}$} & \multicolumn{1}{r}{${\small 0.654}$} & \multicolumn{1}{r}{${\small %
0.039}$} & \multicolumn{1}{r|}{${\small 0.481}$} & \multicolumn{1}{|r}{$%
{\small -0.073}$} & \multicolumn{1}{r}{${\small 0.907}$} &
\multicolumn{1}{r}{${\small -0.372}$} & \multicolumn{1}{r}{${\small 1.130}$}
& ${\small 22.013}$ {\small sec} \\
\multicolumn{1}{l}{${\small 50}$} & \multicolumn{1}{r}{${\small -0.046}$} &
\multicolumn{1}{r}{${\small 0.253}$} & \multicolumn{1}{r}{${\small 0.022}$}
& \multicolumn{1}{r|}{${\small 0.139}$} & \multicolumn{1}{|r}{${\small -0.043%
}$} & \multicolumn{1}{r}{${\small 0.487}$} & \multicolumn{1}{r}{${\small %
0.018}$} & \multicolumn{1}{r|}{${\small 0.367}$} & \multicolumn{1}{|r}{$%
{\small -0.032}$} & \multicolumn{1}{r}{${\small 0.723}$} &
\multicolumn{1}{r}{${\small 0.261}$} & \multicolumn{1}{r}{${\small 0.916}$}
& ${\small 49.563}${\small \ sec} \\
\multicolumn{1}{l}{${\small 100}$} & \multicolumn{1}{r}{${\small -0.026}$} &
\multicolumn{1}{r}{${\small 0.173}$} & \multicolumn{1}{r}{${\small 0.009}$}
& \multicolumn{1}{r|}{${\small 0.097}$} & \multicolumn{1}{|r}{${\small -0.023%
}$} & \multicolumn{1}{r}{${\small 0.350}$} & \multicolumn{1}{r}{${\small %
0.012}$} & \multicolumn{1}{r|}{${\small 0.262}$} & \multicolumn{1}{|r}{$%
{\small -0.027}$} & \multicolumn{1}{r}{${\small 0.548}$} &
\multicolumn{1}{r}{${\small -0.089}$} & \multicolumn{1}{r}{${\small 0.733}$}
& ${\small 2.789}${\small \ mins} \\
${\small 200}$ & \multicolumn{1}{r}{${\small -0.011}$} & \multicolumn{1}{r}{$%
{\small 0.122}$} & \multicolumn{1}{r}{${\small 0.002}$} &
\multicolumn{1}{r|}{${\small 0.064}$} & \multicolumn{1}{|r}{${\small -0.009}$%
} & \multicolumn{1}{r}{${\small 0.243}$} & \multicolumn{1}{r}{${\small 0.006}
$} & \multicolumn{1}{r|}{${\small 0.180}$} & \multicolumn{1}{|r}{${\small %
0.003}$} & \multicolumn{1}{r}{${\small 0.386}$} & \multicolumn{1}{r}{$%
{\small -0.056}$} & \multicolumn{1}{r}{${\small 0.506}$} & ${\small 10.370}$%
{\small \ mins} \\
\multicolumn{1}{l}{${\small 500}$} & \multicolumn{1}{r}{${\small -0.005}$} &
\multicolumn{1}{r}{${\small 0.075}$} & \multicolumn{1}{r}{${\small 0.000}$}
& \multicolumn{1}{r|}{${\small 0.041}$} & \multicolumn{1}{|r}{${\small -0.007%
}$} & \multicolumn{1}{r}{${\small 0.155}$} & \multicolumn{1}{r}{${\small %
0.003}$} & \multicolumn{1}{r|}{${\small 0.117}$} & \multicolumn{1}{|r}{$%
{\small -0.007}$} & \multicolumn{1}{r}{${\small 0.241}$} &
\multicolumn{1}{r}{${\small -0.026}$} & \multicolumn{1}{r}{${\small 0.323}$}
& ${\small 1.035}${\small \ hours} \\ \hline\hline
\end{tabular}
\ \ \ \ ${\small
\caption{Bias and RMSE of CM estimator of
two-parameter transformed Gumbel copula.}}\label{T2}%
\end{table}%

\noindent%
\begin{table}[h] \centering%
$%
\begin{tabular}{ccccccccccccccccc}
& \multicolumn{4}{c}{${\small \tau=0.01}$} & \multicolumn{4}{c}{${\small %
\tau =0.2}$} & \multicolumn{4}{c}{${\small \tau=0.5}$} & \multicolumn{4}{c}{$%
{\small \tau=0.8}$} \\ \hline
& \multicolumn{2}{c}{${\small \alpha=0.1}$} & \multicolumn{2}{c}{${\small %
\beta=1.059}$} & \multicolumn{2}{c}{${\small \alpha=0.2}$} &
\multicolumn{2}{c}{${\small \beta =1.137}$} & \multicolumn{2}{c}{${\small %
\alpha=0.5}$} & \multicolumn{2}{c}{${\small \beta=1.6}$} &
\multicolumn{2}{c}{${\small \alpha =0.9}$} & \multicolumn{2}{c}{${\small %
\beta=3.45}$} \\ \hline
& {\small Bias} & {\small RMSE} & {\small Bias} & {\small RMSE} & {\small %
Bias} & {\small RMSE} & {\small Bias} & {\small RMSE} & {\small Bias} &
{\small RMSE} & {\small Bias} & {\small RMSE} & {\small Bias} & {\small RMSE}
& {\small Bias} & {\small RMSE} \\ \hline\hline
\multicolumn{17}{c}{${\small n=30}$} \\
\multicolumn{1}{l}{\small CM} & \multicolumn{1}{r}{${\small -0.082}$} &
\multicolumn{1}{r}{${\small 0.313}$} & \multicolumn{1}{r}{${\small 0.039}$}
& \multicolumn{1}{r}{${\small 0.190}$} & ${\small -0.064}$ & ${\small 0.642}$
& ${\small 0.033}$ & ${\small 0.486}$ & ${\small -0.068}$ & ${\small 0.567}$
& ${\small 0.074}$ & ${\small 0.466}$ & ${\small -0.072}$ & ${\small 0.955}$
& ${\small -0.359}$ & ${\small 1.128}$ \\
\multicolumn{1}{l}{\small PML} & \multicolumn{1}{r}{${\small -0.067}$} &
\multicolumn{1}{r}{${\small 0.068}$} & \multicolumn{1}{r}{${\small -0.485}$}
& \multicolumn{1}{r}{${\small 0.486}$} & ${\small -0.117}$ & ${\small 0.128}$
& ${\small -0.561}$ & ${\small 0.568}$ & ${\small 0.078}$ & ${\small 0.394}$
& ${\small -0.426}$ & ${\small 0.594}$ & ${\small -0.043}$ & ${\small 0.421}$
& ${\small 0.253}$ & ${\small 1.027}$ \\
\multicolumn{1}{l}{${\small \rho}${\small -}${\small \tau}$} &
\multicolumn{1}{r}{${\small 1.236}$} & \multicolumn{1}{r}{${\small 2.895}$}
& \multicolumn{1}{r}{${\small -0.213}$} & \multicolumn{1}{r}{${\small 1.775}$%
} & ${\small 1.101}$ & ${\small 2.984}$ & ${\small -0.913}$ & ${\small 1.641}
$ & ${\small -0.556}$ & ${\small 2.312}$ & ${\small -0.975}$ & ${\small 1.142%
}$ & ${\small -0.439}$ & ${\small 0.691}$ & ${\small 0.919}$ & ${\small 1.039%
}$ \\ \hline\hline
\multicolumn{17}{c}{${\small n=50}$} \\
\multicolumn{1}{l}{\small CM} & \multicolumn{1}{r}{${\small -0.046}$} &
\multicolumn{1}{r}{${\small 0.245}$} & \multicolumn{1}{r}{${\small -0.021}$}
& \multicolumn{1}{r}{${\small 0.141}$} & ${\small -0.052}$ & ${\small 0.468}$
& ${\small 0.029}$ & ${\small 0.357}$ & ${\small -0.037}$ & ${\small 0.506}$
& ${\small 0.015}$ & ${\small 0.364}$ & ${\small -0.033}$ & ${\small 0.732}$
& ${\small 0.206}$ & ${\small 0.892}$ \\
\multicolumn{1}{l}{\small PML} & \multicolumn{1}{r}{${\small -0.060}$} &
\multicolumn{1}{r}{${\small 0.062}$} & \multicolumn{1}{r}{${\small -0.478}$}
& \multicolumn{1}{r}{${\small 0.482}$} & ${\small 0.102}$ & ${\small 0.112}$
& ${\small -0.516}$ & ${\small 0.526}$ & ${\small -0.072}$ & ${\small 0.240}$
& ${\small -0.472}$ & ${\small 0.556}$ & ${\small 0.025}$ & ${\small 0.315}$
& ${\small 0.330}$ & ${\small 0.772}$ \\
\multicolumn{1}{l}{${\small \rho}${\small -}${\small \tau}$} &
\multicolumn{1}{r}{${\small 1.115}$} & \multicolumn{1}{r}{${\small 2.033}$}
& \multicolumn{1}{r}{${\small -0.289}$} & \multicolumn{1}{r}{${\small 1.378}$%
} & ${\small 1.035}$ & ${\small 2.537}$ & ${\small -0.354}$ & ${\small 1.341}
$ & ${\small -0.478}$ & ${\small 2.110}$ & ${\small -0.952}$ & ${\small 1.021%
}$ & ${\small -0.392}$ & ${\small 0.508}$ & ${\small 0.801}$ & ${\small 0.991%
}$ \\ \hline\hline
\multicolumn{17}{c}{${\small n=100}$} \\
\multicolumn{1}{l}{\small CM} & \multicolumn{1}{r}{${\small -0.022}$} &
\multicolumn{1}{r}{${\small 0.171}$} & \multicolumn{1}{r}{${\small 0.009}$}
& \multicolumn{1}{r}{${\small 0.100}$} & ${\small -0.019}$ & ${\small 0.342}$
& ${\small 0.016}$ & ${\small 0.258}$ & ${\small 0.029}$ & ${\small 0.367}$
& ${\small -0.005}$ & ${\small 0.257}$ & ${\small -0.025}$ & ${\small 0.551}$
& ${\small 0.155}$ & ${\small 0.704}$ \\
\multicolumn{1}{l}{\small PML} & \multicolumn{1}{r}{${\small -0.058}$} &
\multicolumn{1}{r}{${\small 0.059}$} & \multicolumn{1}{r}{${\small -0.483}$}
& \multicolumn{1}{r}{${\small 0.485}$} & ${\small -0.109}$ & ${\small 0.113}$
& ${\small -0.524}$ & ${\small 0.528}$ & ${\small -0.071}$ & ${\small 0.158}$
& ${\small -0.403}$ & ${\small 0.469}$ & ${\small -0.023}$ & ${\small 0.167}$
& ${\small -0.017}$ & ${\small 0.237}$ \\
\multicolumn{1}{l}{${\small \rho}${\small -}${\small \tau}$} &
\multicolumn{1}{r}{${\small 0.973}$} & \multicolumn{1}{r}{${\small 1.220}$}
& \multicolumn{1}{r}{${\small -0.176}$} & \multicolumn{1}{r}{${\small 1.273}$%
} & ${\small 0.923}$ & ${\small 2.335}$ & ${\small -0.340}$ & ${\small 1.457}
$ & ${\small -0.365}$ & ${\small 1.114}$ & ${\small -0.852}$ & ${\small 1.001%
}$ & ${\small -0.255}$ & ${\small 0.397}$ & ${\small 0.708}$ & ${\small 0.686%
}$ \\ \hline\hline
\multicolumn{17}{c}{${\small n=200}$} \\
\multicolumn{1}{l}{\small CM} & \multicolumn{1}{r}{${\small -0.016}$} &
\multicolumn{1}{r}{${\small 0.122}$} & \multicolumn{1}{r}{${\small 0.002}$}
& \multicolumn{1}{r}{${\small 0.064}$} & ${\small -0.011}$ & ${\small 0.244}$
& ${\small 0.001}$ & ${\small 0.383}$ & ${\small -0.014}$ & ${\small 0.245}$
& ${\small 0.005}$ & ${\small 0.180}$ & ${\small -0.001}$ & ${\small 0.396}$
& ${\small -0.060}$ & ${\small 0.527}$ \\
\multicolumn{1}{l}{\small PML} & \multicolumn{1}{r}{${\small -0.041}$} &
\multicolumn{1}{r}{${\small 0.062}$} & \multicolumn{1}{r}{${\small -0.503}$}
& \multicolumn{1}{r}{${\small 0.505}$} & ${\small -0.099}$ & ${\small 0.102}$
& ${\small -0.514}$ & ${\small 0.516}$ & ${\small -0.050}$ & ${\small 0.116}$
& ${\small -0.333}$ & ${\small 0.415}$ & ${\small -0.059}$ & ${\small 0.144}$
& ${\small 0.038}$ & ${\small 0.400}$ \\
\multicolumn{1}{l}{${\small \rho}${\small -}${\small \tau}$} &
\multicolumn{1}{r}{${\small 0.874}$} & \multicolumn{1}{r}{${\small 1.025}$}
& \multicolumn{1}{r}{${\small -0.235}$} & \multicolumn{1}{r}{${\small 1.215}$%
} & $-{\small 0.890}$ & ${\small 2.414}$ & ${\small -0.330}$ & ${\small 1.041%
}$ & ${\small -0.321}$ & ${\small 0.997}$ & ${\small -0.786}$ & ${\small %
0.988}$ & ${\small -0.239}$ & ${\small 0.331}$ & ${\small 0.580}$ & ${\small %
0.625}$ \\ \hline\hline
\end{tabular}
\ \ \ \ \ ${\small
\caption{Bias and RMSE of  CM, PML and ${\tau}$-${\rho}$ estimators of
two-parameter transformed Gumbel copula.}}\label{T3}%
\end{table}%
}
\end{landscape}

\section{\label{section5}Comments and conclusions}

\noindent From Table 2, we conclude that by considering three dependence
cases: weak $\left( \tau =0.01\right) ,$ moderate $\left( \tau =0.5\right) $
and strong $\left( \tau =0.8\right) ,$ the performance, in terms of bias and
RMSE, of the CM based estimation is well justified. In each case, for small
and large samples, the bias and RMSE are sufficiently small.\ Moreover, in
time-consuming point of view, we observe that for a sample size $n=30$ and
for $N=1000$ replications, the central processing unit (CPU) time to process
CM's method took $22.013$ seconds, which is relatively small. For one
replication $N=1,$ the CPU time (in seconds) for different sample sizes are
summarized as follows: $\left( n,CPU\right) =\left( 30,0.213\right) ,$ $%
\left( 100,0.312\right) ,$ $\left( 200,0.844\right) ,$ $\left(
500,3.922\right) .$ Table 3 shows that both the PML and the CM based
estimation perform better than the $\left( \tau ,\rho \right) $-inversion
method. However, in weak dependence case $\tau =0.01{\small ,}$ the CM
method provides better results than the PML one, mainly when the sample size
increases. On the other hand, it is worth mentioning that our method is
quick with respect to the PML one. The main advantage of our method is that
it provides estimators with explicit forms, as far as Archimedean copula
models are concerned. This is not the case of the other methods which
require numerical procedures leading to eventual problems in execution time
and inaccuracy issues. In conclusion, the CM based estimation method
performs well for the chosen model. Furthermore, its usefulness in the weak
dependence case particularly makes it a good candidate for statistical tests
of independence.\bigskip

\noindent \textbf{Acknowledgement. }The authors are indebted to an anonymous
referee for valuable remarks and suggestions.

\appendix

\section{Appendix}

\subsection{\label{ProofTH1}Proof of Theorem \protect\ref{TH1}}

\noindent By considering CM's estimator as a Z-estimator (van der Vaart,
1998, page 41), a straight application of Theorem 1 in Tsukahara (2005)
leads to the consistency and asymptotic normality of the considered
estimator. Indeed, the existence of a sequence of consistent roots $\widehat{%
\mathbf{\theta }}^{CM}$ to $\left( \ref{CM-estimator}\right) ,$ may be
verified by using similar arguments as the proof of Theorem 1 in Tsukahara
(2005). More precisely, we have to check only the conditions in Theorem
A.10.2 in Bickel \textit{et al.} (1993). Indeed, first recall $\left( \ref%
{Lk}\right) $ and set%
\begin{equation*}
\Phi \left( \mathbf{\theta }\right) :=\int_{\mathbb{I}^{d}}\mathbf{L}\left( 
\mathbf{u};\mathbf{\theta }\right) dC_{\mathbf{\theta }_{0}}\left( \mathbf{u}%
\right) ,\text{ and }\Phi _{n}\left( \mathbf{\theta }\right)
:=n^{-1}\sum\limits_{i=1}^{n}\mathbf{L}\left( \widehat{\mathbf{U}}_{i};%
\mathbf{\theta }\right) ,
\end{equation*}%
where $\widehat{\mathbf{U}}_{i}=\left( F_{1n}\left( X_{1i}\right)
,...,F_{dn}\left( X_{di}\right) \right) ,$ with $\left(
X_{j1},...,X_{jn}\right) $ is a given random sample from the rv $X_{j}.$ In
view of assumption $\left[ H.2\right] $ the following derivatives exist%
\begin{equation*}
\overset{\bullet }{\Phi }\left( \mathbf{\theta }\right) =\int_{\mathbb{I}%
^{d}}\overset{\bullet }{\mathbf{L}}\left( \mathbf{u};\mathbf{\theta }\right)
dC_{\mathbf{\theta }_{0}}\left( \mathbf{u}\right) ,\text{ }\overset{\bullet }%
{\Phi }_{n}\left( \mathbf{\theta }\right) =\frac{1}{n}\sum_{i=1}^{n}\overset{%
\bullet }{\mathbf{L}}\left( \widehat{\mathbf{U}}_{i};\mathbf{\theta }\right)
.
\end{equation*}%
Next, we verify that%
\begin{equation}
\sup \left\{ \left\vert \overset{\bullet }{\Phi }_{n}\left( \mathbf{\theta }%
\right) -\overset{\bullet }{\Phi }\left( \mathbf{\theta }\right) \right\vert
:\left\vert \mathbf{\theta }-\mathbf{\theta }_{0}\right\vert <\epsilon
_{n}\right\} \overset{\mathbf{P}}{\rightarrow }0,\text{ as }n\rightarrow
\infty ,  \label{sup}
\end{equation}%
for any real sequence $\epsilon _{n}\rightarrow 0.$ Indeed, since $\overset{%
\bullet }{\mathbf{L}}$ is continuous in $\mathbf{\theta ,}$ then%
\begin{equation*}
\sup \left\{ \left\vert \overset{\bullet }{\mathbf{L}}\left( \widehat{%
\mathbf{U}}_{i};\mathbf{\theta }\right) -\overset{\bullet }{\mathbf{L}}%
\left( \widehat{\mathbf{U}}_{i};\mathbf{\theta }_{0}\right) \right\vert
:\left\vert \mathbf{\theta }-\mathbf{\theta }_{0}\right\vert <\epsilon
_{n}\right\} =o_{\mathbf{P}}\left( 1\right) ,\text{ }i=1,...,n,
\end{equation*}%
and the fact that%
\begin{equation*}
\left\vert \overset{\bullet }{\Phi }_{n}\left( \mathbf{\theta }\right) -%
\overset{\bullet }{\Phi _{n}}\left( \mathbf{\theta }_{0}\right) \right\vert
\leq \frac{1}{n}\sum_{i=1}^{n}\left\vert \overset{\bullet }{\mathbf{L}}%
\left( \widehat{\mathbf{U}}_{i};\mathbf{\theta }\right) -\overset{\bullet }{%
\mathbf{L}}\left( \widehat{\mathbf{U}}_{i};\mathbf{\theta }_{0}\right)
\right\vert .
\end{equation*}%
implies%
\begin{equation}
\sup \left\{ \left\vert \overset{\bullet }{\Phi }_{n}\left( \mathbf{\theta }%
\right) -\overset{\bullet }{\Phi _{n}}\left( \mathbf{\theta }_{0}\right)
\right\vert :\left\vert \mathbf{\theta }-\mathbf{\theta }_{0}\right\vert
<\epsilon _{n}\right\} \overset{\mathbf{P}}{\rightarrow }0,\text{ as }%
n\rightarrow \infty .  \label{sup2}
\end{equation}%
On the other hand, in view of the law of the large number, we have%
\begin{equation*}
\frac{1}{n}\sum_{i=1}^{n}\overset{\bullet }{\mathbf{L}}\left( \mathbf{U}_{i};%
\mathbf{\theta }_{0}\right) \overset{\mathbf{P}}{\rightarrow }\overset{%
\bullet }{\Phi }\left( \mathbf{\theta }_{0}\right) ,\text{ as }n\rightarrow
\infty ,
\end{equation*}%
where $\mathbf{U}_{i}=\left\{ F_{j}\left( X_{ji}\right) \right\} _{1\leq
j\leq d}.$ Moreover, in view of the continuity of function $\overset{\bullet 
}{\mathbf{L}}$ in $\mathbf{u}$ and Glivenko-Cantelli theorem, that is%
\begin{equation*}
\sup_{x_{j}}\left\vert F_{jn}\left( x_{j}\right) -F_{j}\left( x_{j}\right)
\right\vert \rightarrow 0,\text{ }j=1,...,d,\text{ almost surely, as }%
n\rightarrow \infty ,
\end{equation*}%
we have%
\begin{equation*}
n^{-1}\sum_{i=1}^{n}\left\vert \overset{\bullet }{\mathbf{L}}\left( \widehat{%
\mathbf{U}}_{i};\mathbf{\theta }_{0}\right) -\overset{\bullet }{\mathbf{L}}%
\left( \mathbf{U}_{i};\mathbf{\theta }_{0}\right) \right\vert \overset{%
\mathbf{P}}{\rightarrow }0.
\end{equation*}%
It follows that $\left\vert \overset{\bullet }{\Phi }_{n}\left( \mathbf{%
\theta }_{0}\right) -\overset{\bullet }{\Phi }\left( \mathbf{\theta }%
_{0}\right) \right\vert \overset{\mathbf{P}}{\rightarrow }0,$ which together
with (\ref{sup2}), implies (\ref{sup}). Conditions (MG0) and (MG3) in
Theorem A.10.2 in Bickel \textit{et al.} (1993) are trivially satisfied by
our assumptions $\left[ H1\right] -\left[ H3\right] .$ In view of the
general theorem for Z-estimators (see, van der Vaart and Wellner, 1996,
Theorem 3.3.1), it remains to prove that $\sqrt{n}\left( \overset{\bullet }{%
\Phi }_{n}-\overset{\bullet }{\Phi }\right) \left( \mathbf{\theta }%
_{0}\right) $ converges in law to the appropriate limit. But this follows
from Proposition 3 in Tsukahara (2005), which achieves the proof of Theorem %
\ref{TH1}.\hfill $\square $

\subsection{Proof Theorem \protect\ref{TH2}}

The proof of Theorem \ref{TH2} is straightforward by using similar argument
as the proof of Theorem \ref{TH1}, therefore the details are omitted.\hfill $%
\square $

\subsection{\label{Consi}Consistency of $\left( \protect\tau ,\protect\rho %
\right) $-inversion estimators}

\noindent In this section we give assumptions on copula models $C_{\mathbf{%
\theta }},$ satisfying condition (\ref{concordance}), that allow consistency
of $\left( \tau ,\rho \right) $-inversion estimators of $\mathbf{\theta }%
=\left( \theta _{1},\theta _{2}\right) $ defined in $\left( \ref{rho-tau-sys}%
\right) .$ On the other terms, we propose some conditions of copula family $%
C_{\mathbf{\theta }}$ ensuring, for large sample sizes, both existence and
uniqueness of system $\left( \ref{rho-tau-sys}\right) .\ $The idea is to
express $\left( \tau ,\rho \right) $-inversion estimators in terms of
Z-estimators (van der Vaart, 1998, page 41) and then we use similar
assumptions allowing consistency of these estimators (see, van der Vaart and
Wellner, 1996, Theorem 3.3.1). Indeed, recall that Kendall's tau and
Spearman's rho corresponding to the couple of rv's $\left(
X_{1},X_{2}\right) $ of dependence function $C_{\mathbf{\theta }}$ are
defined, respectively, by%
\begin{equation*}
\left\{ 
\begin{array}{l}
\tau \left( \mathbf{\theta }\right) =4\dint_{\left[ 0,1\right] ^{2}}C_{%
\mathbf{\theta }}\left( u_{1},u_{2}\right) dC_{\mathbf{\theta }}\left(
u_{1},u_{2}\right) \mathbf{-}1\medskip \\ 
\rho \left( \mathbf{\theta }\right) =12\dint_{\left[ 0,1\right]
^{2}}u_{1}u_{2}dC_{\mathbf{\theta }}\left( u_{1},u_{2}\right) \mathbf{-}3.%
\end{array}%
\right.
\end{equation*}%
It is easy to verify that (\ref{rho-tau-sys}) is equivalent to the following
system%
\begin{equation}
\left\{ 
\begin{array}{c}
\dsum\limits_{i=1}^{n}L_{\tau }\left( F_{1n}\left( X_{1i}\right)
,F_{2n}\left( X_{2i}\right) ;\mathbf{\theta }\right) =0 \\ 
\dsum\limits_{i=1}^{n}L_{\rho }\left( F_{1n}\left( X_{1i}\right)
,F_{2n}\left( X_{2i}\right) ;\mathbf{\theta }\right) =0,%
\end{array}%
\right.  \label{Raz}
\end{equation}%
where%
\begin{equation*}
L_{\tau }\left( u_{1},u_{2};\mathbf{\theta }\right) :=4C_{\mathbf{\theta }%
}\left( u_{1},u_{2}\right) -1-\tau \left( \mathbf{\theta }\right) ,
\end{equation*}%
and 
\begin{equation*}
L_{\rho }\left( u_{1},u_{2};\mathbf{\theta }\right) :=12u_{1}u_{2}-3-\rho
\left( \mathbf{\theta }\right) ,
\end{equation*}%
with $F_{jn}$ denotes the empirical df pertaining to the sample $\left(
X_{j1},...,X_{jn}\right) $ defined in $\left( \ref{Fn}\right) .$ This
implies, by representation $\left( \ref{Raz}\right) ,$ that $\left( \tau
,\rho \right) $-inversion estimators of the true value $\mathbf{\theta }_{0}$
are, indeed, Z-estimators. Therefore, by using the Z-estimation theory, we
conclude that consistency of such estimators may be established provided
that the following two assumptions hold:

\begin{itemize}
\item $\left[ A.1\right] $ $\mathbf{\theta }_{0},$ element of an open $%
\mathcal{O}\subset \mathbb{R}^{2},$ is the unique zero of the mapping%
\begin{equation*}
\mathbf{\theta }\rightarrow \int_{\left[ 0,1\right] ^{2}}\mathbb{L}\left(
u_{1},u_{2};\mathbf{\theta }\right) dC_{\mathbf{\theta }_{0}}\left(
u_{1},u_{2}\right) ,
\end{equation*}
defined from $\mathcal{O}$ to $\mathbb{R}^{2},$ with $\mathbb{L}\left(
u_{1},u_{2};\mathbf{\theta }\right) :=\left( L_{\tau }\left( u_{1},u_{2};%
\mathbf{\theta }\right) ,L_{\rho }\left( u_{1},u_{2};\mathbf{\theta }\right)
\right) .$

\item $\left[ A.2\right] $ $\mathbb{L}\left( \cdot ;\mathbf{\theta }\right) $
is differentiable with respect to $\mathbf{\theta }$ with the Jacobian
matrix denoted by%
\begin{equation*}
\overset{\bullet }{\mathbb{L}}\left( u_{1},u_{2};\mathbf{\theta }\right) :=%
\left[ 
\begin{tabular}{cc}
$\frac{\partial L_{\tau }\left( u_{1},u_{2};\mathbf{\theta }\right) }{%
\partial \theta _{1}}$ & $\frac{\partial L_{\tau }\left( u_{1},u_{2};\mathbf{%
\theta }\right) }{\partial \theta _{2}}$ \\ 
$\frac{\partial L_{\rho }\left( u_{1},u_{2};\mathbf{\theta }\right) }{%
\partial \theta _{1}}$ & $\frac{\partial L_{\rho }\left( u_{1},u_{2};\mathbf{%
\theta }\right) }{\partial \theta _{2}}$%
\end{tabular}%
\right] ,
\end{equation*}%
$\overset{\bullet }{\mathbb{L}}\left( u_{1},u_{2};\mathbf{\theta }\right) $
is continuous both in $\left( u_{1},u_{2}\right) $ and $\mathbf{\theta },$
and the Euclidian norm $\left\vert \overset{\bullet }{\mathbb{L}}\left(
u_{1},u_{2};\mathbf{\theta }\right) \right\vert $ is dominated by a $dC_{%
\mathbf{\theta }}$-integrable function $g\left( u_{1},u_{2}\right) .$
\end{itemize}


\begin{thebibliography}{99}
\bibitem{} Bickel, P. J., Klaassen, C. A. J., Ritov, Y. and Wellner, J. A.,
1993. Efficient and adaptive estimation for semiparametric models. Johns
Hopkins Series in the Mathematical Sciences. \textit{Johns Hopkins
University Press, Baltimore, MD.}

\bibitem{} Cherbini, U., Luciano, E. and Vecchiato, W., 2004. Copula methods
in finance. \textit{Wiley Finance Series. John Wiley \& Sons, Ltd.,
Chichester.}

\bibitem{} Deheuvels, P., 1979. La fonction de d\'{e}pendance empirique et
ses propri\'{e}t\'{e}s. \textit{Acad. Roy. Belg. Bull. Cl. Sci.} \textbf{65}%
, 274-292.

\bibitem{} Genest, C., 1987. Frank's family of bivariate distributions. 
\textit{Biometrika,} \textbf{74}, 549-555.

\bibitem{} Genest, C., Ghoudi, K. and Rivest, L. P., 1995. A semiparametric
estimation procedure of dependence parameters in multivariate families of
distributions. \textit{Biometrika} \textbf{82}, 543-552.

\bibitem{} Joe, H., 1997. \textit{Multivariate Models and Dependence Concepts%
}, Chapman \& Hall, London.

\bibitem{} Joe, H., 2005. Asymptotic efficiency of the two-stage estimation
method for copula-based models. \textit{J. Multivariate Anal.} \textbf{94},
401-419.

\bibitem{} Kim, G., Silvapulle, M. J. and Silvapulle, P., 2007. Comparison
of semiparametric and parametric methods for estimating copulas. \textit{%
Comm. Statist. Simulation Comput.} \textbf{51}, 2836-2850.

\bibitem{} Nelsen, R. B., 2006. \textit{An Introduction to Copulas}, second
ed. Springer, New York.

\bibitem{} Oakes, D., 1982. A model for association in bivariate survival
data. \textit{J. Roy. Statist. Soc. Ser. B} \textbf{44}, no. 3, 414-422.

\bibitem{} Schmid, F., Schmidt, R., Blumentritt, T., Gai\ss er, S. and
Ruppert, M., 2010. Copula-Based Measures of Multivariate Association. 
\textit{Lecture Notes in Statistics,} \textbf{1}, Volume 198, Copula Theory
and Its Applications, Part 1, Pages 209-236.

\bibitem{} Sklar, A., 1959. Fonctions de r\'{e}partition \`{a} $n$
dimensions et leurs marges, \textit{Publ. Inst. Statist.} Univ. Paris 8,
229-231.

\bibitem{} Tsukahara, H., 2005. Semiparametric estimation in copula models.
Canad. J. Statist. \textbf{33}, 357-375.

\bibitem{} van der Vaart, A. W. and Wellner, J. A., 1996. \textit{Weak
Convergence and Empirical Processes: With applications to Statistics}.
Springer, New York.

\bibitem{} van der Vaart, A. W., 1998. \textit{Asymptotic Statistics},
Cambridge University Press.

\bibitem{} Yan, J., 2007. Enjoy the Joy of Copulas: With a Package copula. 
\textit{Journal of Statistical Software,} \textbf{21}(4), 1-21.
\end{thebibliography}
\end{document}